\documentclass{ws-ijmpa}

\begin{document}

\markboth{Authors' Names}
{Instructions for Typing Manuscripts (Paper's Title)}

%
\catchline{}{}{}{}{}
%

\title{DYNAMICALLY GENERATED RESONANCES IN THE CHIRAL
UNITARY APPROACH TO MESON BARYON INTERACTION}

\author{\footnotesize E.~OSET, S. SARKAR, M.J. VICENTE VACAS}

\address{Departamento de F\'\i sica Te\'orica and IFIC,
Centro Mixto Universidad de Valencia-CSIC,
Institutos de Investigaci\'on de Paterna, Aptd. 22085, 46071
Valencia, Spain
}

\author{A. RAMOS}

\address{Departament d'Estructura i Constituents de la Mat\`eria,
Universitat de Barcelona, \\ 
        Diagonal 647, 08028 Barcelona, Spain
}

\author{D. JIDO}

\address{Physik-Department,  Technische Universit\"at M\"unchen
D-85747 Garching, Germany
}

\author{J.A. OLLER}

\address{Departamento de F\'{\i}sica, Universidad de Murcia, 30071
  Murcia, Spain
}
\author{U.G. MEISSNER}

\address{Universit\"at Bonn, Helmholtz-Institut f\"ur Strahlen- und
Kernphysik (Theorie)\\
 Nu{\ss}alle 14-16, D-53115 Bonn, Germany
}

\maketitle

\pub{Received (Day Month Year)}{Revised (Day Month Year)}

\begin{abstract}
In this talk we report on the use of a chiral unitary approach for the 
 interaction of the octets of meson and baryon and the octet of mesons with the
 decuplet of baryons. Two octets of $J^{\pi}=1/2^-$  baryon states
  and a singlet are generated dynamically in the first case,  resulting in 
  the case of
strangeness $S=-1$ in two poles of the scattering matrix close to the nominal
$\Lambda(1405)$ resonance. In the second case many resonances are also
generated, among them an exotic baryon with $S=1$ corresponding to a $\Delta K$
resonance.  We make suggestions of experiments which could show
evidence for the existence of these states.

\keywords{Keyword1; keyword2; keyword3.}
\end{abstract}
\section{Description of the meson baryon interactions}
The introduction of unitarity constraints in coupled channels in chiral
perturbation theory has led to unitary extensions of the theory that starting
from the same effective Lagrangians allow one to make predictions at much higher
energies . One of the interesting
consequences of these extensions is that they generate dynamically low lying
resonances, both in the mesonic and baryonic sectors. By this we mean that they
are generated by the multiple scattering of the meson or baryon components,
much  the same as the deuteron is generated by the interaction of the nucleons
through the action of a  potential, and they are not preexistent states that
remain in the large $N_c$ limit where the multiple scattering is suppressed.

    Starting from the chiral Lagrangians for the interaction of the octets of 
    meson and baryon 
\cite{chiral} and using the N/D method to obtain a scattering matrix fulfilling
exactly unitarity in coupled channels \cite{oller}, the full set of transition matrix
elements with the coupled channels  is given in matrix form by

\begin{equation}
T = [1 - V \, G]^{-1}\, V~.
\label{eq:bs1}
\end{equation}
Here, the matrix $V$, obtained from the lowest order meson--baryon
chiral Lagrangian, contains the Weinberg-Tomozawa or seagull
contribution, as employed e.g.  in Ref.~\cite{bennhold},

\begin{equation}
V_{i j} = - C_{i j} \frac{1}{4 f^2}(2\sqrt{s} - M_{i}-M_{j})
\left(\frac{M_{i}+E}{2M_{i}}\right)^{1/2} \left(\frac{M_{j}+E^{\prime}}{2M_{j}}
\right)^{1/2}~ ,
\label{eq:ampl2}
\end{equation}
For strangeness $S=-1$ the $C_{i j}$ coefficients are given in Ref.~\cite{angels},
and an averaged meson decay constant $f=1.123f_{\pi}$ is used
\cite{bennhold}, with $f_\pi = 92.4\,$MeV the weak pion decay constant.
 
The diagonal
matrix $G$ stands
for the loop function of a meson and a baryon and is defined by a subtracted
 dispersion
relation in terms of phase space with a cut starting at the corresponding
threshold \cite{oller}. It
corresponds to the loop function of a meson and a baryon once the
logarithmic divergent constant is removed.
The analytical properties of $G$ are properly kept when evaluating
the previous loop function in dimensional regularization and an explicit
subtraction constant,$a_i$, appears in the expression.

This meson baryon loop function  was calculated
in Ref.~\cite{angels} with a cut-off regularization.
%
The values of the $a_{i}$
 constants 
are found to be around $-2$ to agree with the results of
the cut--off method for cut--off values of the order of the mass of
the $\rho(770)$ \cite{oller}, which we call of natural size.

\section{Poles of the T-matrix}
\label{sec:3}

  The study of Ref.~\cite{bennhold} showed the presence of  poles in
  Eq.~(\ref{eq:bs1}) around the $\Lambda(1405)$ and the
$\Lambda(1670)$ for isospin $I=0$ and around the $\Sigma(1620)$ in
$I=1$.  The same approach in $S=-2$ leads to the resonance
$\Xi(1620)$ \cite{xi} and in $S=0$ to the $N^*(1535)$
\cite{siegel,inoue}. 
 One is thus tempted to consider the appearance of a singlet and an octet
of meson--baryon resonances. Nevertheless, the situation is more
complicated because indeed in the SU(3) limit there are {\it two} octets
and not just one, as we discuss below. 
The presence of these multiplets was already
  discussed in Ref.~\cite{oller} after obtaining a pole with $S=-1$ in the
$I=1$ channel,
with mass around 1430 MeV, and two poles with $I=0$, of masses around that of the
  $\Lambda(1405)$.

 The appearance of a multiplet of dynamically generated mesons and baryons seems
 most natural once a state of the multiplet appears. Indeed, one must recall that the
 chiral Lagrangians are obtained from the combination of the octet of
 pseudoscalar mesons (the pions and partners) and the octet of stable baryons
(the nucleons and partners).  The SU(3) decomposition of the combination of two
 octets tells us that
 \begin{equation}
 8 \otimes 8=1\oplus 8_s \oplus 8_a \oplus 10 \oplus \overline{10} \oplus 27~.
\end{equation}
Thus, on pure SU(3) grounds, should we have a SU(3) symmetric Lagrangian,
 one can expect e.g. one singlet and two octets of resonances, the symmetric and
 antisymmetric ones.  

 The lowest order of the meson--baryon chiral Lagrangian is exactly SU(3) 
invariant if
 all the masses of the mesons, or equivalently the quark
 masses,  are set equal. As stated above  [see Eq.~(\ref{eq:ampl2})], 
in Ref.~\cite{bennhold}
 the baryon masses take their physical values, although strictly
 speaking at the leading order in the chiral expansion they should be equal to
 $M_0$. For Eq.~(\ref{eq:ampl2}) being SU(3) symmetric, all the baryons masses
 $M_{i}$ must be set equal as well. When all the meson and baryon masses are
  equal, and these common masses are employed in evaluating the $G_l$ functions,
 together with equal subtraction constants $a_l$, the $T$--matrix obtained
from Eq.~(\ref{eq:bs1}) is also SU(3) symmetric. 

If we do such an SU(3) symmetry approximation
and look for poles of the scattering matrix, we find poles
corresponding to the octets and singlet. The surprising result is that
the two octet poles are degenerate as a consequence of the 
 dynamics contained in
 the chiral Lagrangians. Indeed, if we evaluate the matrix elements of the transition potential
$V$ in a basis of SU(3) states, we obtain something proportional to 
$V_{\alpha  \beta}= {\rm diag}(6,3,3,0,0,-2)$
taking the following order for the irreducible representations:
$1$, $8_s$, $8_a$, $10$, $\overline{10}$ and $27$, with positive sign meaning
attraction.

Hence we observe that the states belonging to different
irreducible representations do not mix and the two octets appear
degenerate. The coefficients in $V_{\alpha  \beta}$
 clearly illustrate why there are no bound
states in the $10$, $\overline{10}$ and $27$ representations.

In practice, the same chiral Lagrangians allow for SU(3) breaking. In
the case of Refs.~\cite{angels,bennhold} the breaking appears
because both in the $V_{i j}$ transition potentials as in the $G_l$
loop functions one uses the
physical masses of the particles as well as different subtraction constants in $G_l$, 
corresponding to the use of a unique cut-off in all channels. 
 In Ref.~\cite{oller} the
physical masses are also used in the $G_l$ functions, although these functions are evaluated 
with a unique subtraction constant as corresponds to the SU(3) limit. 
In both approaches, physical masses are
used to evaluate the $G_l$ loop functions so that unitarity is fulfilled
exactly and the physical thresholds for all channels are respected. 

\begin{figure}
\centerline{\psfig{file=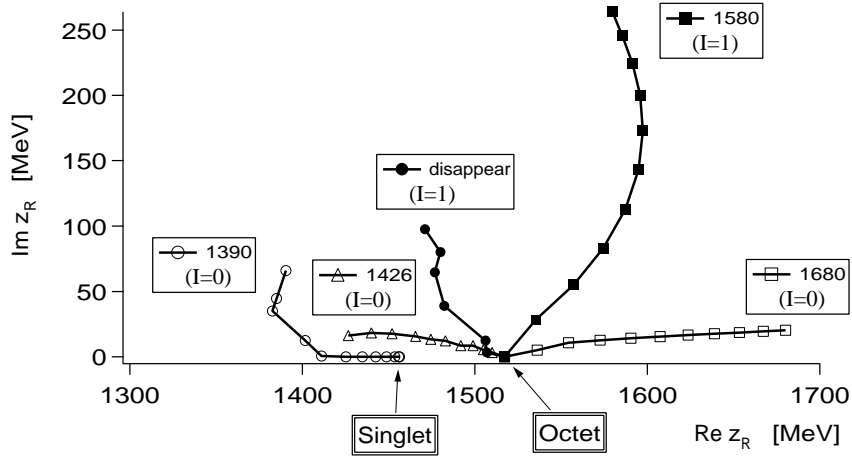,width=12cm}}
\vspace*{8pt}
  \caption{Trajectories of the poles in the scattering amplitudes obtained by
  changing the SU(3) breaking parameter $x$ gradually. At the SU(3) symmetric 
  limit ($x=0$),
   only two poles appear, one is for the singlet and the other for the octet.
  The symbols correspond to the step size $\delta x =0.1$. The results are from
  \protect\cite{Jido:2003cb}.}
  \label{fig:tracepole}
\end{figure}

By following the approach of Ref.~\cite{bennhold} and using the
physical masses of the baryons and the mesons, the position of the
poles change and the two octets split apart in four branches, two
for $I=0$ and two for $I=1$, as one can see in
Fig.~\ref{fig:tracepole}. In the figure we show the trajectories
of the poles as a function of a parameter $x$ that breaks
gradually the SU(3) symmetry up to the physical values.  The
dependence of masses and subtraction constants on the parameter
$x$ is given by
\begin{eqnarray}
M_i(x) &= & M_0+x(M_i-M_0),  \nonumber \\
m^{2}_{i}(x) &=& m_{0}^{2} + x (m^{2}_{i}-m^{2}_{0}), \nonumber\\
a_{i}(x) &=& a_{0} + x (a_{i} - a_{0}),
\end{eqnarray}
where $0\le x \le 1$.  In the calculation of
Fig.~\ref{fig:tracepole}, the values $M_{0}=1151$ MeV, $m_{0} =
368$ MeV and $a_{0}= -2.148 $ are used.

  The complex poles, $z_R$, appear in unphysical sheets. In the
present search we follow the strategy of changing
the sign of the momentum $q_l$ 
in the $G_l(z)$ loop function  for the channels which
are open at an energy equal to Re($z$).

The splitting of the two $I=0$ octet states is very interesting.
One moves to higher energies to merge with the $\Lambda(1670)$
resonance and the other one moves to lower energies to create a
pole, quite well identified below the  $\bar{K}N$ threshold, with
a narrow width. 
 On the other hand, the singlet also evolves
to produce a pole at low energies with a quite large width.

We note that the singlet and the $I=0$ octet states appear
nearby in energy and  what experiments  actually
see is a combination of the effect of these two resonances.

Similarly as for the $I=0$ octet states, we can see that one branch of the 
$I=1$ states moves to higher energies while another
moves to lower energies. The branch moving to higher energies finishes at
what would correspond to the $\Sigma(1620)$ resonance when the physical
masses are reached. The
branch moving to lower energies fades away after a while when getting close to
the $\bar{K}N$ threshold.  

The model of Ref.~\cite{oller} reproduces qualitatively the same results, but
the $I=1$ pole also stays for $x=1$. Nevertheless,
 in both approaches there
is an $I=1$ amplitude with an enhanced strength around the $\bar{K} N$ threshold.
 This amplitude has non negligible  consequences for 
reactions producing $\pi \Sigma$ pairs in that region.
This has been illustrated for instance
in Ref.~\cite{nacher1}, where the photoproduction of the $\Lambda(1405)$ via the
reaction $\gamma p \to K^+ \Lambda(1405)$ was studied. It was shown there that
the different sign in the
$I=1$ component of the $\mid \pi^+ \Sigma^-\rangle$, $\mid \pi^- \Sigma^+\rangle$
states leads, through interference between the $I=1$ and the dominant $I=0$
amplitudes, to different
cross sections in the various charge channels, a fact that has been
confirmed experimentally very recently \cite{ahn}.

Once the pole positions are found, one can also determine the
couplings of these resonances to the physical states by studying
the amplitudes close to the pole and identifying them with
\begin{equation}
T_{i j}=\frac{g_i g_j}{z-z_R}~.
\end{equation}
The couplings $g_i$ are in general complex valued numbers.
In the Table  we summarize the
pole positions and the complex couplings $g_i$ obtained from the
model of Ref.~\cite{bennhold} for isospin $I=0$. The results with the model of
\cite{oller} are qualitatively similar.

\begin{table}[ht]
\centering 
\caption{\small Pole positions and couplings to $I=0$
physical states from the model of Ref.~\protect\cite{bennhold}}
 \vspace{0.5cm}
\begin{tabular}{|c|cc|cc|cc|}
\hline
 $z_{R}$ & \multicolumn{2}{c|}{$1390 + 66i$} &
\multicolumn{2}{c|}{$1426 + 16i$} &
 \multicolumn{2}{c|}{$1680 + 20i$}  \\
 $(I=0)$ & $g_i$ & $|g_i|$ & $g_i$ & $|g_i|$ & $g_i$ & $|g_i|$ \\
 \hline
 $\pi \Sigma$ & $-2.5-1.5i$ & 2.9 & $0.42-1.4i$ & 1.5 & $-0.003-0.27i$ &
 0.27 \\
 ${\bar K} N$ & $1.2+1.7i$ & 2.1 & $-2.5+0.94i$ & 2.7 & $0.30+0.71i$ &
 0.77 \\
 $\eta\Lambda$ & $0.010+0.77i$ & 0.77 & $-1.4+0.21i$ & 1.4 & $-1.1-0.12i$ &
 1.1 \\
 $K\Xi$ & $-0.45-0.41i$ & 0.61 & $0.11-0.33i$ & 0.35 & $3.4+0.14i$ &
 3.5 \\
 \hline
 \end{tabular}
\label{tab:jido0}
\end{table}

 We observe that the
second resonance with $I=0$ couples strongly to $\bar{K} N$ channel, while
the first resonance couples more strongly to $\pi \Sigma$.

\section{Influence of the poles on the physical observables}
\label{sec:5}

In a given reaction the $\Lambda(1405)$ resonance is always seen in $\pi \Sigma$
mass distribution. However, the $\Lambda(1405)$ can be produced through any of
the channels in the Table. Hence,
it is clear that,  should there be a reaction which forces this
initial channel to be $\bar{K}N$, then this would give more
weight to the second resonance, $R_{2}$, and hence produce a
distribution with a shape corresponding to an effective resonance
narrower than the nominal one and at higher energy. Such a case
indeed occurs in the reaction $K^- p \to \Lambda(1405) \gamma$
studied theoretically in Ref.~\cite{nacher}.  It was shown there
that since the $K^- p$ system has a larger energy than the
resonance, one has to lose energy emitting a photon prior to the
creation of the resonance and this is effectively done by the
Bremsstrahlung from the original $K^-$ or the proton.  Hence the
resonance is initiated from the  $K^- p$ channel. This is also the case
in the reaction $\gamma p\to K^* \Lambda(1405)$ which has also the 
$\Lambda(1405)$ initiated by $\bar{K} N$ through the vertex 
$\gamma \to K^* K$ \cite{Hyodo:2004vt}.

\section{The interaction of the decuplet of baryons with the octet of mesons}
Given the success of the chiral unitary approach in generating dynamically low
energy resonances from the interaction of the octets of stable baryons
and the pseudoscalar mesons,   in 
\cite{Kolomeitsev:2003kt} the
interaction of the decuplet of $3/2^+$ with the octet of pseudoscalar mesons 
was studied and shown to
lead to many states which were associated to experimentally well 
established $3/2^-$ resonances. 
  
    The lowest order chiral Lagrangian for the interaction of the baryon 
decuplet with the octet of pseudoscalar mesons is given by \cite{Jenkins:1991es}
\begin{equation}
L=i\bar T^\mu D_{\nu} \gamma^{\nu} T_\mu -m_T\bar T^\mu T_\mu
\label{lag1} 
\end{equation}
where $T^\mu_{abc}$ is the spin decuplet field and $D^{\nu}$ the covariant derivative
given by in \cite{Jenkins:1991es}. The identification of the physical decuplet
states with the $T^\mu_{abc}$ can be seen in \cite{sarkar}, where a detailed
study of this interaction and the resonances generated can be seen. The study is
done along the lines of the former sections, looking for poles in the second
Riemann sheet of the complex plane, the coupling of the resonances to the
different channels and the stability of the results with respect to variations
of the input parameter, which in our case is just the subtraction constant,a.
This allows the association of the resonances found to existing states of the
particle data book, and the prediction of new ones.  A detail of the results
obtained can be seen in Fig. 2.

  Another interesting result is the generation of an exotic state of $S=1$ 
  and $I=1$ which  is generated  by the interaction of the $\Delta K $ channels
  and stands as a $\Delta K$ resonance. The pole appears in a Riemann sheet
  below threshold when also the sign of the momentum is changed, but it leads
  to a $\Delta K $ amplitude which accumulates strength close to threshold and 
produces a broad peak in the cross section, see Fig.3,  in contrast to the
$I=2$ cross section which is much smaller and very smooth.

In order to investigate this interaction we propose the study 
of the
following reactions:  1) $pp \to \Lambda \Delta^+ K^+$, 2) $pp \to \Sigma^- \Delta^{++}
K^+$, 3) $pp \to \Sigma^0 \Delta^{++}K^0$. In the first case the $\Delta^+ K^+$ state
produced has necessarily $I=1$.  In the second case the $\Delta^{++}K^+$ state has
$I=2$. In the third case the $\Delta^{++}K^0$ state has mostly an $I=1$ component.
The study of the
$\Delta K$ invariant mass distribution in these reactions would provide
experimental information to eventually prove the prediction made here, and thus
giving support to this new exotic baryonic state, which, although in the quark
model would require at least five quarks and could qualify as a pentaquark, 
finds a much simpler interpretation as a
resonant $\Delta K$ state. 

\begin{figure}
\centerline{\psfig{file=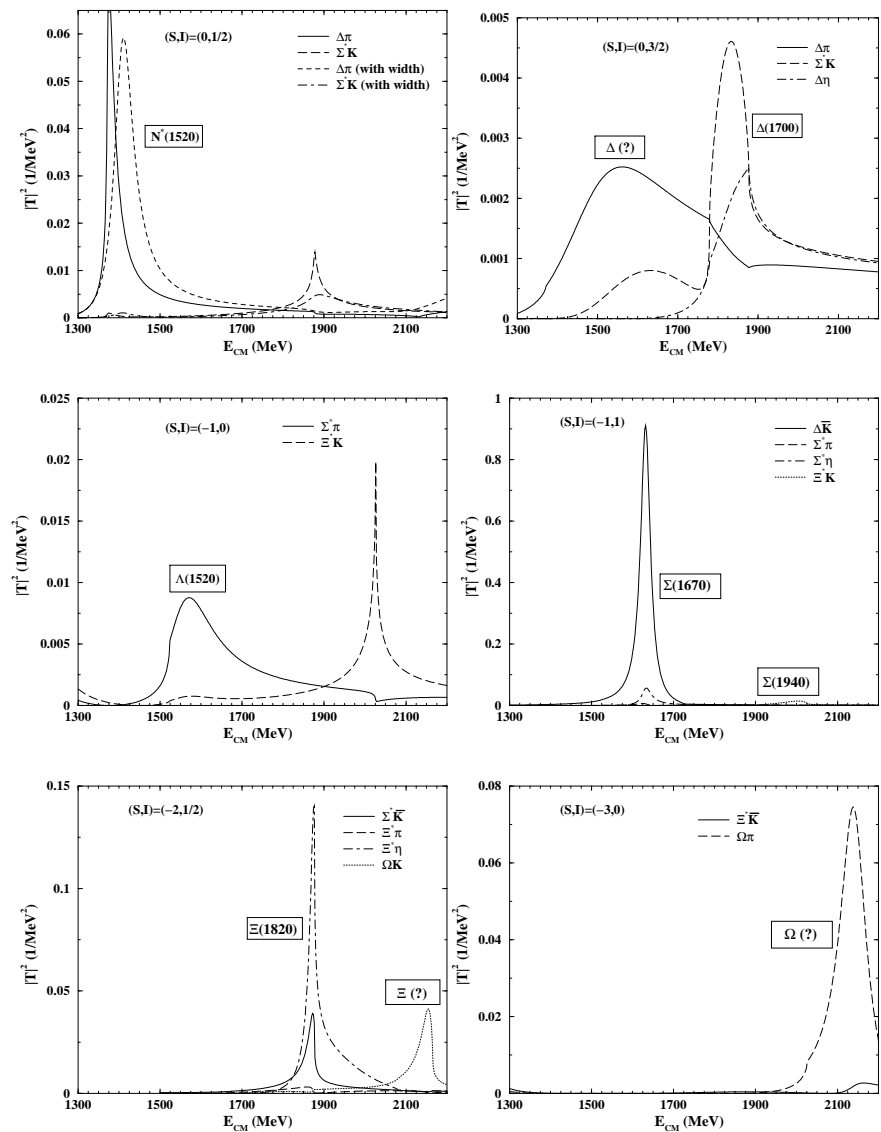,width=14.5cm}}
\caption{Resonances obtained from the interaction of the octet of mesons with
  the decuplet of baryons.}
  \label{fig:decu}
\end{figure}

\begin{figure}[tbp]
    \centering
    \includegraphics[width=10cm,clip]{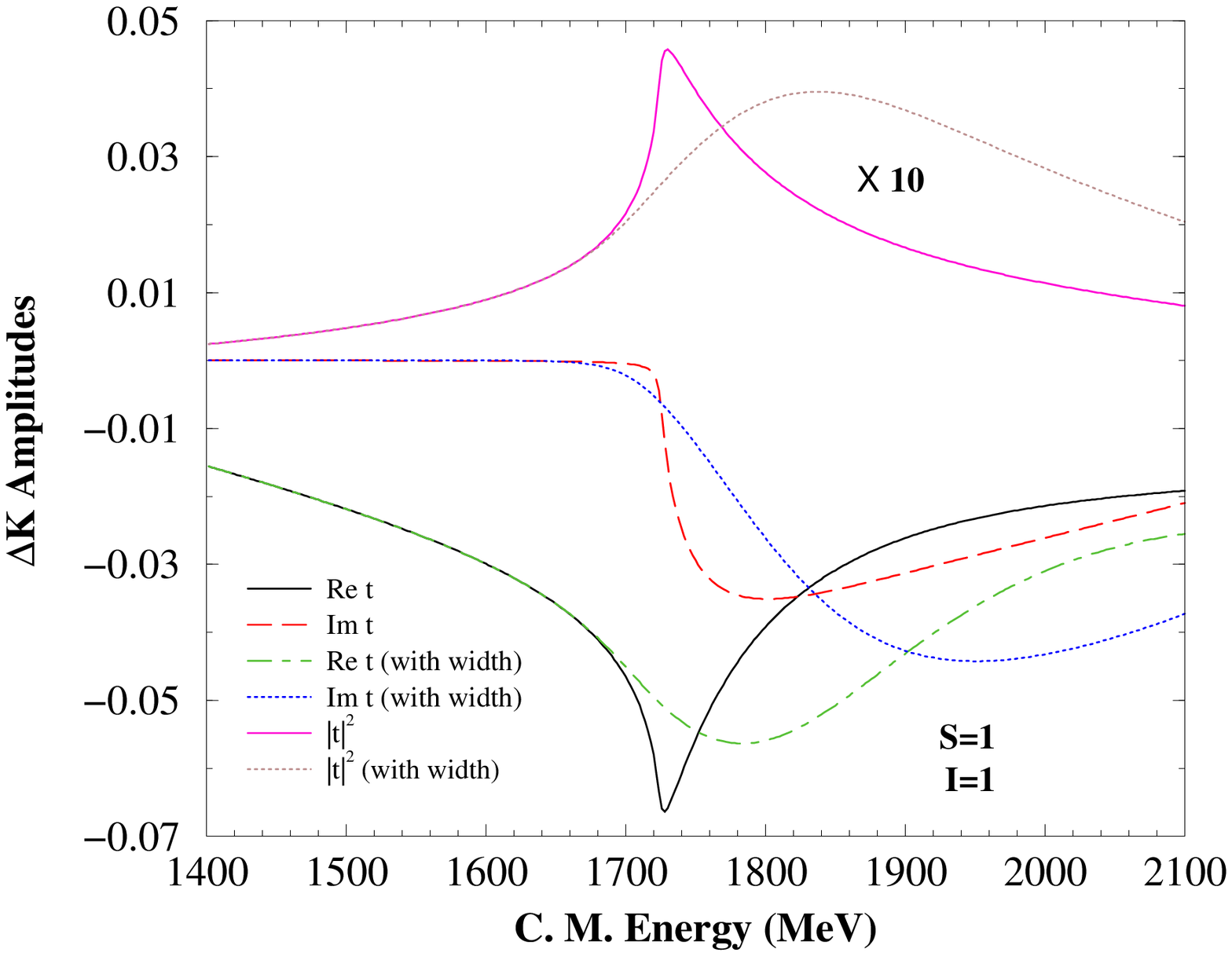}   
\caption{Amplitudes for $\Delta K\rightarrow\Delta K$ for $I=1$}
\end{figure}

\section{Acknowledgments}
This work is partly supported by DGICYT contract number BFM2003-00856, 
CICYT  (Spain)  Grant Nos. FPA2002-03265 and FPA2004-03470, 
 the E.U. EURIDICE network contract no. HPRN-CT-2002-00311 and the Research
 Cooperation program of the  JSPS and the  CSIC.

\end{document}